\documentclass[letterpaper]{IEEEtran}
\IEEEoverridecommandlockouts
\usepackage{cite}
\usepackage{amsmath,amssymb,amsfonts}
\usepackage{algorithmic}
\usepackage{graphicx}
\usepackage{textcomp}
\usepackage{xcolor}
\def\BibTeX{{\rm B\kern-.05em{\sc i\kern-.025em b}\kern-.08em
    T\kern-.1667em\lower.7ex\hbox{E}\kern-.125emX}}
\begin{document}

\title{Distributed Learning for Vehicular Dynamic Spectrum Access in Autonomous Driving \\
\thanks{Copyright © 2022 IEEE. Personal use is permitted. For any other purposes, permission must be obtained from the IEEE by emailing pubspermissions@ieee.org. This is the author’s version of an article that has been accepted for publication in the proceedings of The 20th International Conference on Pervasive Computing and Communications (PerCom 2022), 4th International Workshop on Pervasive Computing for Vehicular Systems (PerVehicle 2022) and will be published by IEEE.\\
The work has been realized within project no. 2018/29/B/ST7/01241 funded by the National Science Centre in Poland.}
}

\author{\IEEEauthorblockN{Pawe\l~Sroka}
\IEEEauthorblockA{\textit{Institute of Radiocommunications} \\
\textit{Poznan University of Technology}\\
Poznań, Poland \\
pawel.sroka@put.poznan.pl; ORCID:0000-0003-0553-7088 \\}
\and
\IEEEauthorblockN{Adrian Kliks}
\IEEEauthorblockA{\textit{Institute of Radiocommunications} \\
\textit{Poznan University of Technology}\\
Poznań, Poland \\
adrian.kliks@put.poznan.pl; ORCID: 0000-0001-6766-7836 \\}
}

\maketitle
\thispagestyle{empty}
\begin{abstract}
Reliable wireless communication between the autonomously driving cars is one of the fundamental needs for guaranteeing passenger safety and comfort. However, when the number of communicating cars increases, the transmission quality may be significantly degraded due to too high occupancy radio of the used frequency band. In this paper, we concentrate on the autonomous vehicle-platooning use-case, where intra-platoon communication is done in the dynamically selected frequency band, other than nominally devoted for such purposes. The carrier selection  is done in a flexible manner with the support of the context database located at the roadside unit (edge of wireless communication infrastructure). However, as the database delivers only context information to the platoons' leaders, the final decision is made separately by the individual platoons, following the suggestions made by the artificial intelligence algorithms. In this work, we concentrate on a lightweight Q-learning solution, that could be successfully implemented in each car for dynamic channel selection. 
\end{abstract}


\section{Introduction}
\label{sec:intro}
The level of advancement of contemporary cars is rising continuously, making car driving more and more safer and comfortable for the passengers. Various functions, which quite recently have not been available or were operated manually by the car driver, are now controlled by the on-board computer. All these achievements bring us closer to the realization of the idea of fully autonomous cars, whose price will be acceptable to the broad community. However, full implementation of this concept entails the need for equipping each car with technologically advanced solutions from the domain of artificial intelligence, allowing cars for making reliable decisions, and thus - for safe self-driving.   \\
In that context, vehicles platooning belongs to one of the key and most promising use cases in the domain of autonomous driving. In such a situation, groups of vehicles drive one-by-one keeping possibly short inter-car distances to reduce fuel consumption and maximize road capacity. Such a convoy of cars can be formed either statically (i.e. in advance, when the group of cars is known) or dynamically (where single cars can join or leave the convoy). The platoon is typically led by a leader which is responsible for sending steering information to other platoon members.\\
As shown in e.g. \cite{b1}, to guarantee safety on the road while reducing inter-car distances (and in consequence carbon footprint) highly reliable intra-platoon communications are obligatory. In that context, one may identify two main approaches towards the vehicle-to-vehicle, V2V (or in a broader sense vehicle-to-anything, V2X) in wireless communication standards. The first option assumes the so-called dedicated short-range communications (DSRC), and the second - usage of cellular networks (cellular-V2X, C-V2X). The former relies on the IEEE 802.11p and wireless access in vehicular environment (WAVE) standards.
Both solutions are known to work sufficiently in normal mode and for currently considered services. However, the increasing ubiquity and pervasiveness of cars across the world generate new challenges here. As it was shown in \cite{b2}, \cite{b3}, DSRC will suffer from medium congestion in case of high traffic load on the roads, which may clearly contribute to the increased number of car accidents. It is then justified to offload a portion of the whole traffic to other, non-congested bands. Various bands could be considered following the findings from the spectrum occupancy measurement campaigns, such as the unoccupied television channels (widely known as the TV white spaces, TVWS \cite{b4}). This approach follows the concept of vehicular dynamic spectrum access (VDSA) discussed in e.g. \cite{b5}, being the extension of the dynamic spectrum access (DSA) scheme explored widely in the cognitive radio (CR) domain \cite{b6}. The idea of application of DSA in vehicular scenarios have been evaluated in various contexts: with the application of artificial intelligence tools \cite{b5}, \cite{b7}, using the queuing theory as in \cite{b9}, \cite{b10}, or even by mimicking the behavior of bumblebee species \cite{b11}, \cite{b12}.\\
Following our prior work \cite{b13zb16,b17zb14}, we focus here on the TV band where autonomous platooning is treated as a secondary service.  In order to protect the primary service (i.e. digital terrestrial television, DTT) and improve the VDSA procedure,  we apply a dedicated database called radio environment map (REM) or context database (CDB) \cite{b15}. As both the location and configuration of the DTT towers, as well as the frequency allocation plan for TV broadcasting services are rather stable in time, such information may be stored in REMs. Moreover, such data may be complemented with high precision measurements of the received power measured across the road. Thus, secondary services may effectively utilize the unoccupied TV bands while protecting DTT receivers due to the accurate data saved in the CDB.\\
In this paper, we investigate further the idea of the usage of CDB and artificial intelligence tools for increasing the VDSA reliability in the selection of appropriate TV band for V2V data transmission. However, as in \cite{wwrf} the focus was on a fully centralized scenario, where the edge intelligence concept was utilized to provide the best frequency channel selection \cite{b16zb13}, in this work we address a fully distributed case. By assumption, the CDB is responsible only for delivering the context information (i.e., the DTT and other platoons' locations) to all platoons (cars) in the service area, and the final learning process and the whole decision making are done at each car or platoon leader. This approach reflects the pervasiveness features, meaning that such a procedure may be realized by each car, as the selected AI tool is intentionally kept at the lowest possible complexity level.  In particular, we have proposed the application of the Q-learning scheme applied in a distributed way to help platoons in the autonomous selection of the best set of communication channels.\\ 
The remainder of the paper is arranged as follows. In the next section, we briefly present the system model and discuss the channel selection problem. Next, in Sec. 3 we present the applied distributed Q-learning-based VDSA. It is followed by the analysis of the achieved simulation results. The whole paper is then concluded. 

\section{System model}
\label{sec:system_model}
\begin{figure}[!htb]
\includegraphics[width=0.5\textwidth]{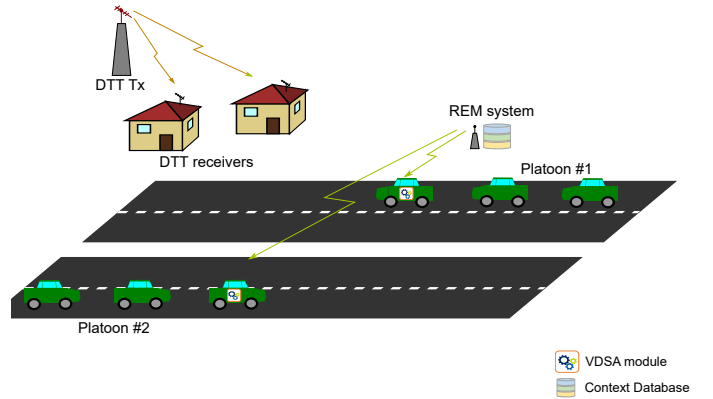}
\caption{Illustration of the conisdered scenario with REM-aided intra-platoon communications in TVWS.}
\label{fig_scenario}
\end{figure}

In this work, we analyze a platooning scenario where the messages related to convoy management are offloaded to the frequency bands other than the 5.9~GHz one dedicated to V2V communications. We consider that the offloaded traffic will be transmitted over the TV band of width $B$ MHz, which is typically split into $K$ equally-sized TV channels, each of either 6 MHz (as typical in the USA) or 8 MHz (as set in e.g. Europe) bandwidth. Without loss of generality, we fix the TV channel width to 8 MHz. Next, we assume that two outermost channels are occupied by the DTT signals originating from the nearby DTT transmitters. The remaining part of the considered frequency band is split into $K_{\mathrm{VDSA}}$ adjacent (partially overlapping) communication channels, each of 10 MHz, as such bandwidth is typical for V2X using the IEEE 802.11p communication standard. Being the primary service, the nearby DTT receivers have to be protected from any harmful interference. Thus, the association of the vacant spectrum to the platoons has to take into consideration this interference aspect. As typically the DTT receivers are mounted permanently at certain locations, it is justified to assume that the location of most of the DTT receivers is fixed.\\
The considered scenario is shown in Fig. \ref{fig_scenario}. It represents a fragment of a highway or high-speed road of length $D$ comprising $L$ lanes. Jointly with separate cars (regular cars or trucks), $M$ platoons may travel at the same time in a short vicinity. The platoon members communicate  periodically to manage the vehicles’ movement via the TV band due to the expected congestion of the 5.9 GHz band. In consequence, up to $M$ frequency channels might need to be allocated to these platoons. As the orthogonal channel assignment among the platoons would be the most preferred option, it is often not possible. In such a case, the algorithm has to propose such allocation pattern that maximizes the communication reliability within the platoon and protects the existing DTT transmission. The distributed dynamic channel selection is supported with context information on the observed DTT signal power along the motorway, acquired from CDB, and on the decisions of other platoons.

\section{Q-LEARNING BASED VDSA ALGORITHM}
\label{sec:qlerning}
As stated before, we propose the application of the reinforcement learning approach, with the Q-learning algorithm employed to identify the best prospective frequency bands for platoons \cite{b14zb17}. This classic AI tool is deployed at each platoon leader and makes use of the information acquired from REM about the expected DTT power for the current location and about the last known position and the selected channel of other platoons. Each platoon may choose an action representing transmission in one of $K_{\mathrm{VDSA}}$ available bands. Thus the total number of possible actions is $K_{\mathrm{VDSA}}$. We create a Q-table representing the aggregate rewards obtained by a platoon for each action depending on the observed platoon state. The state is represented using a tuple comprising information on the expected transmission conditions in each of the available bands. In particular, these conditions are expressed in terms of quantized signal to interference and noise ratio, SINR, assuming $R$ quantization levels. Moreover, the SINR values are sorted in ascending order. The total number of possible states can be then formulated as:
\begin{equation}
    S= { {R+K_{\mathrm{VDSA}}-1} \choose K_{\mathrm{VDSA}}},
\end{equation}
where ${x \choose y} = \frac {x!}{y!(x-y)!}$ denotes the binomial coefficient (choose), with $x!$ representing the factorial of $x$. Therefore, $S$ is equal to the number of possible combinations with repetition when choosing $K_{\mathrm{VDSA}}$ values from a set of $R$ values of quantized SINR. 

For each available channel $k=1, 2,\ldots, K_{\mathrm{VDSA}}$, and for each platoon vehicle $v$ the expected SINR is calculated using the information obtained from REM on the following parameters:
\begin{itemize}
    \item the center frequency, 
    \item the maximum allowed transmit power: $P_{max}^{(k)}(v)$,
    \item the expected interference from DTT system in band $k$ for vehicle $v$: $P_{DTT}^{(k)}(v)$
    \item the expected interference from other platoons in band $k$ for vehicle $v$: $P_{other}^{(k)}(v)$.
\end{itemize}
The state calculation of a platoon $m_v$ for each band $k$ is determined based on the minimum SINR value estimated within the platoon: $min_{v \in m_v} \mathrm{SINR}^{(k)}(v)$, with $\mathrm{SINR}^{(k)}(v)$ calculated as:
\begin{equation}
\label{eq:sinr_v}
    SINR^{(k)}(v) = \min\{\mathrm{SINR}_{0}^{(k)}(v), \mathrm{SINR}_{v-1}^{(k)}(v)\}
\end{equation}
with 
\begin{equation}
\label{eq:sinr_vu}
    SINR_{u}^{(k)}(v) = \frac{P_{max}^{(k)}(u) \times h_{u,v}^{(k)}}{\sigma^{2}+P_{DTT}^{(k)}(v)+P_{other}^{(k)}(v)},
\end{equation}
where $u=0$ describes the platoon leader and $u=v-1$ represents the preceding vehicle, $\sigma^{2}$ is the noise variance, and $h_{u,v}^{(k)}$ denotes the channel gain between vehicles $u$ and $v$, estimated using a predefined path-loss model depending on the distance information acquired from REM. The interference from other platoons is calculated using:
\begin{equation}
\label{eq:p_other}
    P_{other}^{(k)}(v) = \sum_{m_u \neq m_v} \sum_{u \in m_u} {w_{u,v} \times P_{max}^{(l)}(u) \times f^{ACIR}_{l,k} \times h_{u,v}^{(k)}} ,
\end{equation}
where $m_u$ is a platoon different than the one of vehicle $v$, $l$ is the frequency band used by platoon $m_u$,  $w_{u,v}$ is the weighting factor corresponding to the probability that $u$ transmits when $v$ is receiving and $f^{ACIR}_{l,k} \in <0,1>$ is the adjacent channel interference ratio factor characterizing the signal leakage between different frequency bands $l$ and $k$.\\
The optimization following the Q-learning procedure relies on the reward $r$ represented as the estimated platoon average throughput, with the individual throughput for each vehicle $v$ calculated using the truncated Shannon formula:
\begin{equation}
\label{eq:shannon}
    T_{v} = \min(100.0, \sum_{n \in N_{\mathrm{VDSA}}} B \cdot \log_{2}(1+\mathrm{SINR}_{u}^{(k)}(v,n))),
\end{equation}
where $B$ is the used bandwidth, $n$ is the packet index out of the all $N_{\mathrm{VDSA}}$ packets received by vehicle $v$ during a single VDSA period, and $u=\{0, v-1\}$.\\
The reward is then used to update the Q-table as follows
\begin{align}
	Q^{(t+1)}&(s_t,a_t )=Q^t(s_t,a_t)+\\\nonumber
	&+\alpha(r_t-\gamma \max_a \left(Q^t s_{(t+1)},a)\right)-Q^t (s_t,a_t)),
\end{align}
where $s_t$ and $a_t$ denote the state and action in time $t$, respectively, $\alpha$ is the learning rate and $\gamma$ is the discount factor~\cite{sutton18}. Two action selection approaches are considered:
\begin{itemize}
    \item the $\epsilon$-greedy method, where in each algorithm iteration an exploration mode can be used with probability $\epsilon$, with the action selected randomly assuming equal probability of each option, or the greedy approach can be employed with probability $1-\epsilon$, where the action is selected as $\max_a\left(Q^t (s_t,a)\right)$.
    \item the softmax method, where in each algorithm iteration the action is selected according to its probability calculated as:
    \begin{equation}
\label{eq:softmax}
    \mathcal{P}_{s_t,a} = \frac{Q_t(a)/\tau}{\sum_{i=1}^{K_{\mathrm{VDSA}}} Q_t(a_i)/\tau},
\end{equation}
where $\tau$ is the temperature parameter ($\tau \in [1,\infty)$) depending on the number of reward samples collected so far for state $s_t$. This approach allows to balance between the need for exploration in case few samples were collected and exploitation in case sufficient accuracy of Q values has been already obtained.
\end{itemize}
As the Q-learning procedure is performed in a distributed way, the need to aggregate the learned Q-tables arises. Therefore two approaches were considered: one, where the ideal fusion is assumed, which is equivalent to all platoons using the same Q-table, and the other, based on the federated learning approach, where the Q-table is combined using the averaging method.

\section{Simulation results}
\label{sec:simres}
\subsection{Simulation setup}
To evaluate the performance of the proposed reinforcement learning approach, we apply a C++-based simulator to model the motorway traffic, which was developed and verified in [14], [15]. We assume intra-platoon communications with a messaging rate of 5 Hz in the TVWS, employing wireless communications based on the IEEE 802.11p protocol. We preform distributed VDSA using three 10 MHz TVWS bands with center frequencies at: 498~MHz, 506~MHz, and 514~MHz, respectively. These are surrounded by two active 8 MHz DTT channels at 490~MHz and 522~MHz. We consider the transmit power of platoon vehicles to be constrained, following the primary user protection need. Thus, we apply power control, following the procedure presented in [16], effectively limiting the interference introduced to the DTT system. The remaining simulation parameters are summarized in Table \ref{TABLEI}.

\begin{table}[!hbtp]
\centering
\caption{Simulation parameters}
\label{TABLEI}
\begin{tabular}{l|c}
\hline
\textbf{Parameters}  &  \textbf{Values}   \\ \hline
Motorway length                        &  5~km \\ \hline
Number of platoons                        &  \{3, 4\} \\ \hline
Cars in platoon                        &  \{6, 10\} \\ \hline
Inter-car spacing in platoon          &  5~m \\ \hline
CACC message periodicity            &  200 ms   \\ \hline
CACC message size                   &  300~B        \\ \hline
Active DTT frequency bands          &  \{490, 522\}~MHz       \\ \hline
Frequency bands for VDSA        &  \{498, 506, 514\}~MHz    \\ \hline
VDSA procedure periodicity                            & 1~s \\ \hline
Q-learning learning rate $\alpha$                           & 0.1 \\ \hline
Q-learning discount factor $\gamma$                           & 0.7 \\ \hline
Number of protected DTT receivers                   & 10 \\ \hline
Minimum protected DTT power $\Gamma_{PU}$                   & -80~dBm \\ \hline
Minimum required DTT SIR $SIR_{min}^{PU}$  & 39.5~dBm \\ \hline
Latency of acquiring information      & 1~s\\ \hline
Number of simulation runs per point                      &    200 \\ \hline
Single simulation run duration       &     140~s\\ \hline
\end{tabular}
\end{table}

We compare the performance of the proposed Q-learning-based VDSA algorithm in several variants with a conventional distributed approach described in [16], where each platoon individually selects the utilized TVWS band based on the (delayed) information on the decisions of other platoons. To facilitate the proper operation of the Q-learning-based approach, we consider an initial offline training phase with the exploration probability of $\epsilon=1$. We investigate the following configurations of the Q-learning method:
\begin{itemize}
    \item $\epsilon$-greedy Q-learning with ideal fusion (combining) of the learning results of all platoons (which effectively makes all platoons use the same Q-table).
    \item $\epsilon$-greedy Q-learning in a federated learning form, where the individual Q-tables of every platoon are combined at the end of each simulation run using averaging.
    \item softmax-based Q-learning, where the probability of selecting each action is calculated according to the softmax formula (\ref{eq:softmax}), and ideal fusion of Q-tables is performed. 
\end{itemize}
For the evaluation in simulations, we considered the exploration probability $\epsilon=0.01$ for the $\epsilon$-greedy approaches, with the Q-table initially trained using over 1 million reward values. With the softmax-based action selection, the temperature parameter $\tau$ value depends on the number of collected samples for the current state, with values in the range [1,$\infty$) possible, where $\infty$ is used (all actions equally probable) when less than 1000 samples are available, and 1 taken (highest Q action selected) when there are more than 100~000 samples. Two learning stages were considered for softmax: with 300~000 and over 1 million reward samples used to build the Q-table, to represent the scenarios with higher and lower exploration probabilities, respectively.
The simulation results are compared using the following factors:
\begin{itemize}
    \item Estimated probability of successful reception of leaders’ packets – calculated as the ratio of successfully received leader messages to the number of all leader transmission attempts; we consider that all platoon vehicles need to receive messages from the platoon leader and the preceding car – we analyze only the performance of the leader-to-member link as it is typically the weaker one than the connection with the preceding car (due to higher distance).
    \item Average number of frequency changes per simulation run – each change of band requires dissemination of management messages (assumed perfect and instantaneous in this work) and reconfiguration of the radio equipment of platoon vehicles, thus introducing additional cost; in consequence, the lower the number of changes the better.
\end{itemize}
Furthermore, we also investigate the cumulative distribution of the signal-to-interference ratio (SIR) observed at the protected DTT receivers in the two TV bands, to ensure the primary system protection constraint is fulfilled. We assume that the DTT system, as the primary one, should not experience lower SIR values than 39.5 dB if the signal is detectable at the receiver site (i.e., its power is greater than -80 dBm).
\subsection{Evaluation results}
This section summarizes the results obtained in simulations considering two scenarios:
\begin{itemize}
    \item With 3 platoons, each of 6 vehicles, which corresponds to a generally less challenging interference scenario, as the distance between the leader and the last platoon vehicle is approximately 50~m. Here the DTT system will be the main interference source.
    \item With 4 platoons, each comprising 10 vehicles, where the communications between the leader and the last platoon vehicle become very challenging due to the significant distance between the leader and last car (around 80~m) and the high inter-platoon interference. Here, the interference from other platoons will be dominant.
\end{itemize}
\subsubsection{Scenario with 3 platoons of 6 vehicles each}
\begin{figure}[!htb]
\includegraphics[width=0.48\textwidth]{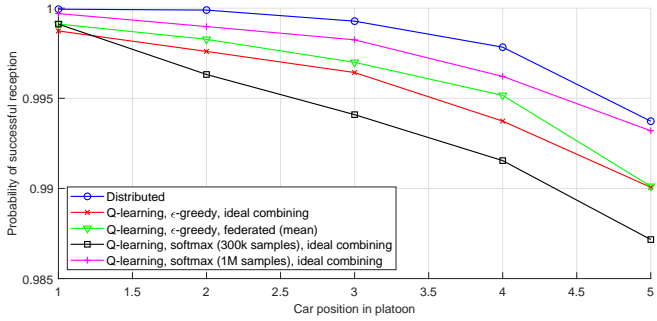}
\caption{Reception rate from leading car vs. position in platoon - scenario with 3 platoons (note that the results are discrete and represented by markers, the lines are introduced only to show the trend).}
\label{fig_Rec_3P}
\end{figure}

\begin{figure}[!htb]
\includegraphics[width=0.48\textwidth]{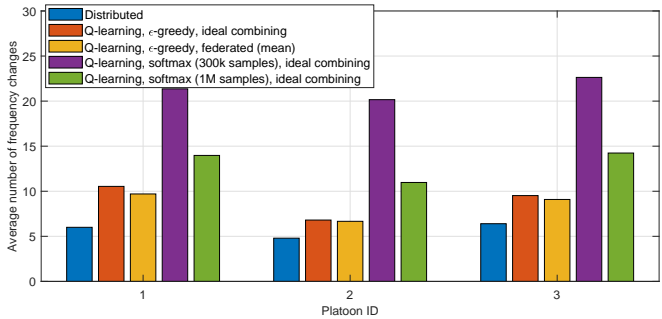}
\caption{Average number of frequency changes per platoon.}
\label{fig_Freq_3P}
\end{figure}

\begin{figure}[!htb]
\includegraphics[width=0.48\textwidth]{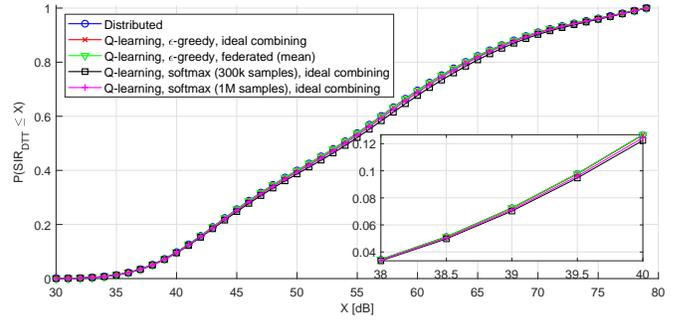}
\caption{SIR cumulative distribution obtained in simulations for the primary system receivers that required protection - channel 522 MHz, 3 platoons.}
\label{fig_SIR_3P}
\end{figure}

Fig. \ref{fig_Rec_3P} presents the estimated probability of successful reception of leader's packets as a function of the position of a vehicle within the platoon, calculated as a ratio of correctly received and decoded leader's messages to the total number of packets transmitted by the leader. One can notice that the best performance is achieved with the conventional distributed approach. Such a result is an expected one, as the interference from other platoons plays a lesser role than the one from the DTT system. The results obtained with the Q-learning method perform slightly worse, which can be the result of the possible (although unlikely) option of exploration, where non-optimal actions are selected. A significant dependence of the probability of successful reception on the number of samples used to initially train the Q-table can be also observed in the case of the softmax approach. This indicates that for the steady-state VDSA system operation the exploration option results in a drop in the reliability of the system. On the other hand, when the Q-table is sufficiently trained and the exploration option is very unlikely, the softmax-based Q-learning method (purple curve) achieves performance that is very close to the one for the distributed algorithm.
\\
In Fig. \ref{fig_Freq_3P} the average number of frequency changes per each of the considered platoons is shown. One can clearly notice the distinctively highest number of band switches in the case of Q-learning with softmax criterion, which was trained with only 300~000 samples. This is a result of the significantly higher probability of exploration in this case than for other configurations. For the remaining Q-learning setups the number of changes is similar, with slightly higher values observed for the softmax criterion. Furthermore, the number of band switches for all Q-learning configurations is higher than for the distributed conventional algorithm, which can be a result of the possibility of exploration mode. This indicates that the cost of VDSA with the use of Q-learning is slightly higher here than for the conventional approach.
\\
Finally, we verified the performance of VDSA in terms of protection of the primary system, with the results of the empirical cumulative distribution of DTT SIR values observed at the receivers shown in Fig. \ref{fig_SIR_3P} (note that we show only the DTT SIR for the 522~MHz band as the conclusions for the 490~MHz band are identical). For all the considered VDSA approaches the performance is almost identical, with less than 10\% of DTT receptions experiencing lower SIR than the one required to fulfill the protection constraint. Such a result was expected as all algorithms employ the same power control method. The weakest 10\% of SIRs, falling below the required threshold, are observed due to the shadowing effect in signal propagation between the platoon and the DTT receiver, resulting in different attenuation of the signal than the one estimated in the power control procedure. Such an effect can be avoided by extending the power control scheme with the use of an additional SIR margin.

\subsubsection{Scenario with 4 platoons of 10 vehicles each}
\begin{figure}[!htb]
\includegraphics[width=0.48\textwidth]{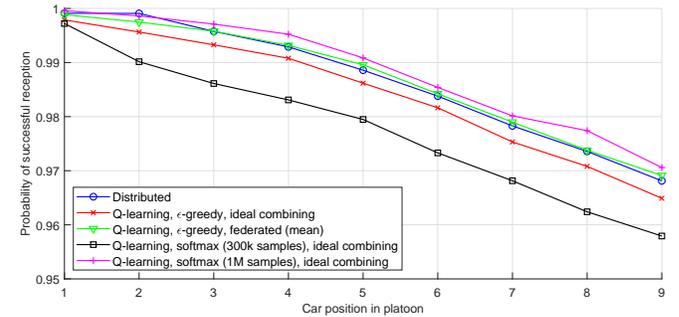}
\caption{Reception rate from leading car vs. position in platoon - scenario with 4 platoons (note that the results are discrete and represented by markers, the lines are introduced only to show the trend).}
\label{fig_Rec_4P}
\end{figure}

\begin{figure}[!htb]
\includegraphics[width=0.48\textwidth]{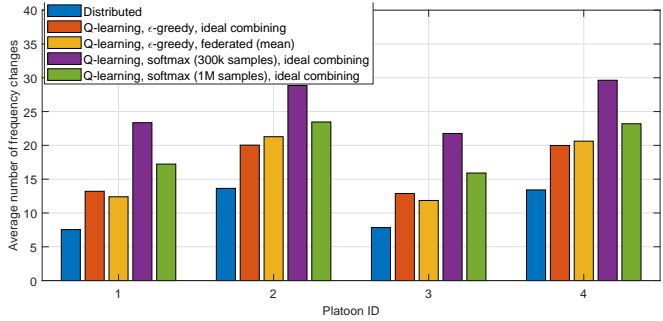}
\caption{Average number of frequency changes per platoon - scenario with 4 platoons.}
\label{fig_Freq_4P}
\end{figure}

\begin{figure}[!htb]
\includegraphics[width=0.48\textwidth]{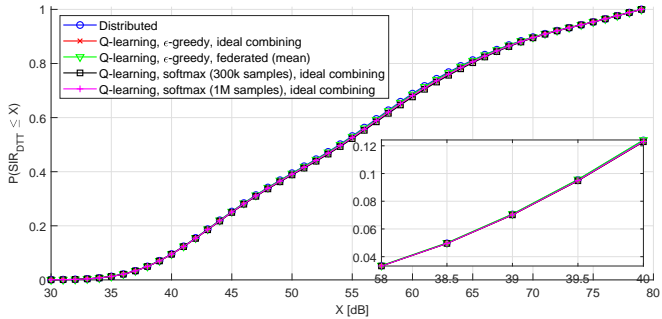}
\caption{SIR cumulative distribution obtained in simulations for the primary system receivers that required protection - channel 522 MHz, 4 platoons.}
\label{fig_SIR_4P}
\end{figure}

Similar simulations were conducted for the scenario with 4 platoons, with the estimated probability of the leader's packets' successful reception shown in Fig. \ref{fig_Rec_4P}. Here, the observed outcome is different than in the case of 3 platoons, as the inter-platoon interference has a significantly higher impact on the performance of intra-platoon communications. One can observe that with 4 platoons the highest reception ratio is obtained with the Q-learning with softmax criterion, which uses the Q-table trained initially with over 1 million samples, which results in a low probability of exploration for most of the possible states. The distributed approach performs slightly worse, as it is unable to fully cope with the inter-platoon interference due to the delay in the exchange of information between the platoons. The performance of $\epsilon$-greedy Q-learning is similar to the one for the distributed algorithm when federated learning is used, while, surprisingly, slightly worse for the ideal fusion. Finally, when using the Q-table trained with fewer samples (300~000) the performance of Q-learning with softmax is significantly worse, as the probability of using exploration mode is higher.
\\
The average number of performed band changes, presented in Fig. \ref{fig_Freq_4P}, shows similar behavior as in the case of 3 platoons. Clearly, the lowest number of switches is performed with the distributed approach, with the values for Q-learning depending on the probability of using the exploration mode. This indicates that the use of Q-learning may be slightly more costly.
\\
When analyzing the protection level of the primary system, with the DTT SIR empirical cumulative distribution shown in Fig. \ref{fig_SIR_4P}, one can conclude that the constraint is fulfilled, with a similar protection level for all considered VDSA algorithms. Again, such an outcome was expected as the same power control scheme is used in all cases.
\subsubsection{Discussion of results}
With the analysis of the estimated probability of successful reception of leader's packets, one can clearly notice that Q-learning algorithms may perform better than the conventional distributed algorithm in case the inter-platoon interference is significant. On the other hand, the exploration mode of Q-learning negatively impacts the reception ratio, as non-optimal actions may be selected. Therefore, an offline training of the Q-table is desirable, e.g. performed when the platoon cars are controlled by human drivers, in order to provide satisfactory reliability for the autonomous platooning. Furthermore, the imperfect combining of Q-tables trained by individual platoons does not seem to negatively impact the outcomes, as the federated learning using averaging provided similar performance to the ideal fusion configuration.\\
The use of Q-learning results in a slightly higher number of band changes than the conventional algorithm, thus increasing the cost of VDSA. This results mostly from the exploration option of Q-learning, however, additional countermeasures, e.g. in form of accounting for the cost of band change in the action selection criterion, should be considered to lower the number of necessary frequency changes.\\
Finally, we can state that the proposed VDSA approach is able to sufficiently protect the primary system. The 10\% of DTT receptions that are below the required threshold result from the shadowing effect, thus can be accounted for with the power control procedure by introducing an additional SIR margin.

\section{Conclusions}
\label{sec:concl}
In this paper, we investigated the application of reinforcement learning-based  VDSA to support the autonomous driving of platoons of vehicles. In particular, the Q-learning approach using different configurations was proposed and tested in two scenarios – with three and four platoons selecting dynamically the communications channel. It has been proved that the proper selection of the strategies and actions in the Q-learning approach led to the achievement of promising results, which may outperform the fully distributed approach. The achieved results confirm the correctness of the proposed methods, and the value of the context information made available by the REM system. However, more intensive investigation towards the action selection criterion with Q-learning is necessary, as frequent band changes were observed. Furthermore, offline training for the Q-learning methods seems necessary, as the operation in exploration mode results in a significant reduction of intra-platoon communications reliability.

\end{document}